\documentclass[conference]{IEEEtran}
\usepackage{graphicx}
\usepackage{hyperref}
\usepackage{booktabs}
\usepackage{amsmath}
\usepackage{listings}
\usepackage{xcolor}
\usepackage{caption}
\usepackage{subcaption}
\usepackage{float}
\usepackage{cite}
\usepackage{fontawesome5}
\usepackage[table]{xcolor}
\usepackage{booktabs}
\usepackage{multirow}
\usepackage{makecell}
\usepackage{graphicx}
\usepackage{tikz}
\usepackage{xcolor}
\usetikzlibrary{calc}
\usepackage{fontawesome5}
\usepackage{mdframed}
\usepackage{caption}
\newmdenv[
  topline=false,
  bottomline=false,
  rightline=false,
  leftline=true,
  linewidth=0.8pt,
  linecolor=black,
  innertopmargin=2pt,
  innerbottommargin=2pt,
  innerleftmargin=8pt,
  innerrightmargin=0pt,
  skipabove=6pt,
  skipbelow=6pt
]{rqanswer}

\definecolor{rqBlack}{HTML}{1A1A1A}
\definecolor{rqDark}{HTML}{4D4D4D}
\definecolor{rqMid}{HTML}{8F8F8F}
\definecolor{rqLight}{HTML}{D4D4D4}
\definecolor{rqGrid}{HTML}{E8E8E8}

\title{REFINE: A Multi-Agent LLM Approach for Evidence-Guided Code Refactoring}
\author{
\IEEEauthorblockN{
Muhammad Waseem\IEEEauthorrefmark{1},
Aakash Ahmad\IEEEauthorrefmark{2}, and
Pekka Abrahamsson\IEEEauthorrefmark{1}
}
\IEEEauthorblockA{
\IEEEauthorrefmark{1}
Faculty of Information Technology and Communication Sciences,\\
Tampere University, Tampere, Finland\\
\{muhammad.waseem, pekka.abrahamsson\}@tuni.fi
}
\IEEEauthorblockA{
\IEEEauthorrefmark{2}
Computing, University of Derby, Derby, United Kingdom\\
a.abbasi@derby.ac.uk
}
}
\begin{document}

\maketitle

\begin{abstract}
Large Language Models have demonstrated strong capabilities in code
generation and transformation, creating new opportunities for automated
software refactoring. However, refactoring requires more than producing
modified code: generated changes should reduce targeted quality problems,
avoid introducing new ones, and preserve behaviour-relevant code structures.
In this paper, we introduce \textbf{REFINE} (\textbf{\underline{R}}efactoring with 
\textbf{\underline{E}}vidence-aware \textbf{\underline{F}}low for 
\textbf{\underline{I}}ntegrated age\textbf{\underline{N}}tic 
\textbf{\underline{E}}xecution), a tool-agnostic evidence-aware multi-agent
approach for Java file-level refactoring candidates. REFINE combines
static-analysis-guided smell identification, smell-informed planning,
LLM-based transformation, automated re-analysis, preservation checks, and
structured reporting.

We empirically evaluate REFINE through a working implementation on 450 Java
files from 15 open-source systems, producing 1,350 model-pass outputs across
three LLM configurations: OpenAI GPT-5.5, Google Gemini 3.1 Pro Preview, and
Anthropic Claude Opus 4.8. The results show substantial reductions in detected
code smells, with total reductions of 68.26\%, 72.79\%, and 68.49\% across the
three configurations, and the strongest reductions observed for major code
smells. A matched 150-file direct-prompt baseline further shows that REFINE
achieves higher median total code-smell reduction with smaller edits and fewer
public-method removals. However, improvements in broader quality indicators
are not uniform: some metrics improve for specific models, while others remain
unchanged or move in an adverse direction. Preservation checks also reveal
residual risks, including critical assert/fail-call changes and public-method
removal. Overall, the results suggest that evidence-aware multi-agent LLM
refactoring is promising for targeted file-level code-smell mitigation, but
neither direct prompting nor the multi-agent workflow provides sufficient
evidence of full behaviour preservation. The generated outputs should
therefore be treated as refactoring candidates that require compilation,
testing, dependency analysis, and human review before adoption in
dependency-rich repository- or system-level settings.
\end{abstract}
\section{Introduction}
\label{sec:introduction}

Software systems continuously evolve to meet changing user needs, fix defects,
add functionality, and adapt to shifting technical and business
requirements~\cite{godfrey2008past}. Over time, these changes can introduce
code smells, degrade maintainability, and increase the need for
refactoring~\cite{lacerda2020code}. Refactoring improves the internal structure
of code while preserving its external behaviour~\cite{fowler2018refactoring}.
However, effective refactoring requires more than applying behaviour-preserving
transformations, and developers must identify relevant quality problems, select
suitable refactoring actions, apply changes consistently, and verify that the
resulting code remains correct and maintainable~\cite{mens2004survey}.

Recent advances in Large Language Models (LLMs) have created new opportunities
for automating software engineering tasks that depend on code understanding,
generation, and transformation~\cite{fan2023large}. In refactoring, these
capabilities are attractive because LLMs can interpret code-quality problems
and generate candidate transformations. However, LLM-assisted refactoring also
introduces practical risks. For example, an LLM may simplify a long method or
split a large class, but the same edit may remove assertions, change public
methods, alter exception-handling logic, or affect framework-sensitive control
flow. Such outcomes may reduce visible code smells while still introducing
behavioural or integration risks. Therefore, LLM-based refactoring should be
evaluated not only by whether code is modified, but also by whether targeted
code smells are reduced, broader quality indicators improve, and
behaviour-relevant structures are preserved.

Existing refactoring research has progressed from rule- and metric-based
automation toward AI-enabled and agentic support. Earlier work emphasized
tool-supported transformations, refactoring opportunity identification, and
quality metrics~\cite{mens2004survey,abid2020thirty}. More recent studies have
explored data-driven refactoring decision support~\cite{martinez2025slr},
foundation-model-based refactoring recommendation~\cite{simoes2025refmodel},
LLM-based code-quality improvement~\cite{alomari2025llms}, and multi-agent
refactoring workflows~\cite{oueslati2025refagent}. These studies show the
potential of LLMs and agents for refactoring, but also highlight open
challenges such as hallucinated suggestions, inconsistent transformations,
limited verification, and difficulty preserving design or architectural
consistency~\cite{alomari2025llms,peitek2026readability,kim2025comparative}.
What remains less clear is how a tool-agnostic evidence-aware multi-agent approach behaves
when its outputs are examined beyond code-smell reduction alone, including
quality movement, preservation risks, failure diagnostics, and model-specific
refactoring behaviour.

In this paper, we introduce \textsc{REFINE}, an evidence-aware multi-agent
workflow for Java file-level refactoring candidates. \textsc{REFINE} combines
static-analysis-guided smell identification, smell-informed planning,
LLM-based transformation, automated re-analysis, preservation checks, and
structured reporting. Detected code smells guide planning, the LLM generates a
candidate transformation, and verification gates record whether the candidate
satisfies configured evidence checks. This design makes each generated
candidate traceable to the code smells that motivated it, the checks it passed
or failed, and the quality or preservation risks observed after transformation.

We evaluate \textsc{REFINE} on 450 Java files from 15 open-source systems,
producing 1,350 model-pass outputs across three LLM configurations. The study
examines file-level code-smell reduction, automated quality indicators, static
preservation proxies, failure diagnostics, and model-specific refactoring
behaviour. To contextualize whether the workflow adds value beyond directly
prompting the same LLMs, we also compare \textsc{REFINE} with a matched
direct-prompt baseline on a 150-file subset. The evaluation shows that
\textsc{REFINE} substantially reduces detected code smells, while broader
quality improvements remain metric- and model-dependent and preservation risks
remain. The matched baseline further suggests that \textsc{REFINE} provides
more controlled file-level candidate generation than direct prompting, although
neither direct prompting nor the multi-agent workflow provides sufficient
evidence of full behaviour preservation. Our goal is therefore not to replace
behaviour-preserving refactoring tools, but to characterize how evidence-aware
LLM workflows generate refactoring candidates, what code-smell reductions they achieve, and what preservation risks remain.

\textbf{Contributions.} This paper makes the following contributions:
\begin{itemize}
\item We introduce \textsc{REFINE}, a tool-agnostic evidence-aware multi-agent
approach for generating Java file-level refactoring candidates, together with a
working implementation~\cite{REFINEAno}.

\item We evaluate \textsc{REFINE} on 450 Java files from 15 systems, producing
1,350 model-pass outputs across three LLM configurations and analysing
code-smell reduction, quality indicators, preservation proxies, failure
diagnostics, and model-specific behaviour.

\item We compare \textsc{REFINE} with direct-prompt baseline on a
150-file subset, evaluating smell reduction, preservation indicators,
public-method removal, and change footprint.

\item We release the experimental dataset~\cite{REFINEDataset}, including
selected files, before--after outputs, model metadata, verification traces,
quality and smell measurements, baseline data, and analysis scripts.
\end{itemize}
\textbf{Implications:}
\textsc{REFINE} provides a traceable way to study LLM-generated Java
refactoring candidates while making clear that such candidates require
additional validation before dependency-rich or system-level adoption.

\section{Approach}
\label{sec:approach}

In this section, we present \textbf{REFINE}, an evidence-aware multi-agent workflow for
generating and evaluating Java file-level refactoring candidates. We first
summarize the workflow shown in Figure~\ref{fig:refine-overview}, and then
describe the implementation, task characterization, orchestration, and
verification trace.

\subsection{Overview}
\label{sec:refine-overview}

\begin{figure*}
    \centering
    \includegraphics[width=0.72\linewidth]{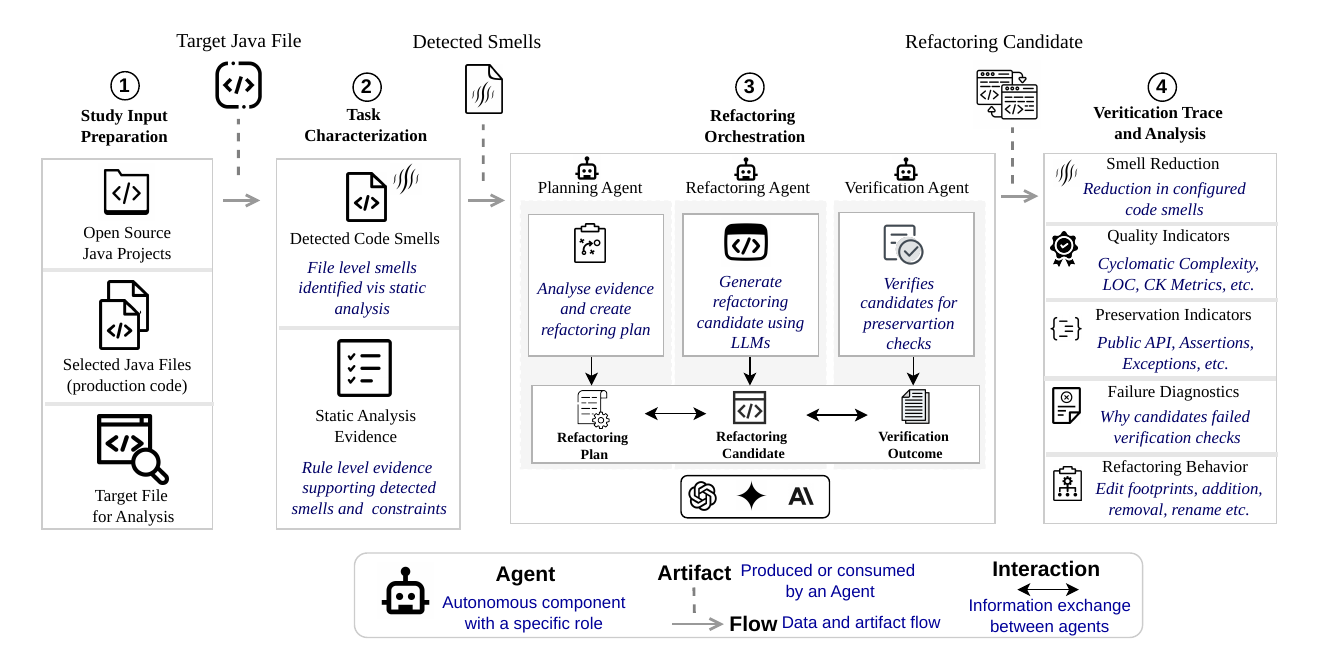}
    \caption{Overview of \textsc{REFINE} for Java file-level refactoring-candidate generation, verification, and trace-based analysis.}
\label{fig:refine-overview}
\end{figure*}

Figure~\ref{fig:refine-overview} summarizes the REFINE workflow used in this
study. In the empirical pipeline, selected Java files from open-source projects
provide the inputs to REFINE. REFINE performs \emph{smell detection} using
static analysis to identify file-level code smells and rule-level evidence.
These detected smells provide the input evidence for the \emph{agent-driven
refactoring} stage, where agents plan the refactoring, invoke an LLM-based
refactoring component, and verify the generated candidate. The resulting
candidate, verification evidence, and intermediate artifacts are then passed to
the \emph{experimental analysis} stage, where we measure code-smell reduction,
quality indicators, structural changes, preservation risks, and refactoring
behaviour.

Internally, \textsc{REFINE} organizes this pipeline around three responsibilities,
namely task characterization, refactoring orchestration, and verification trace
collection. During characterization, \textsc{REFINE} links the target Java file
with detected code smells, refactoring instructions, public API information, and
workspace context. During orchestration, static and LLM-based agents prepare a
refactoring plan, generate a transformation, and verify the candidate. During
trace collection, \textsc{REFINE} persists the transformed source, before--after
measurements, verification outcomes, model metadata, and diagnostics.

The workflow is therefore not a single direct-prompt code rewrite. Detected
code smells guide planning, the LLM generates a transformation, and verification
gates record whether the output satisfies configured evidence checks. Generated
code is treated as a refactoring candidate rather than as an automatically
accepted result.


\subsection{Tool Implementation}
\label{sec:tool-implementation}

\textsc{REFINE} is implemented as a working research prototype with a Next.js
interface, a Java Spring Boot backend for workspace management and static
analysis, and a Python agent service for orchestration, LLM routing,
verification, and metric computation. The Python service uses LangGraph~v1.1 to
coordinate the workflow, while the Java backend invokes PMD~7.x to detect and
re-count code smells before and after transformation. The system supports ZIP
upload, Git repository cloning, and inline Java input, and can run in both
interactive and batch-evaluation modes.

The empirical study uses an artifacts-only configuration in which generated
candidates, verification outcomes, model metadata, before--after metrics,
diagnostics, and reports are saved for analysis, while candidate files are not
written back to the original repositories. This separates controlled candidate
evaluation from project integration and supports reproducibility.

\subsection{Task Characterization}
\label{sec:refine-task}

REFINE forms a bounded refactoring task from a persisted workspace. The input
can originate from a ZIP upload, a Git repository clone, or inline Java source
content. A task is defined at the level of a single Java file and is represented
by the workspace identifier, file path, source content, selected static-analysis
evidence, and protocol metadata such as a frozen sample identifier or
batch-mode flag. This file-level formulation keeps the transformation scope
explicit while preserving the surrounding workspace context.

REFINE characterizes the target file using static evidence and preservation
constraints. In the empirical study, the configured backend is PMD, but the
workflow is not tied to a specific static-analysis tool. Detected code smells
are treated as initial refactoring evidence and prioritized before prompt
construction. The refactoring prompt receives the original source code, a
prioritized subset of detected code smells, rule-based refactoring instructions,
public API constraints, and retry notes when available. REFINE also extracts
public API signatures to identify interface elements that should be preserved
during transformation.

\subsection{Refactoring Orchestration}
\label{sec:refine-orchestration}

The orchestration stage coordinates workflow roles for planning, candidate
generation, verification, and reporting. REFINE registers eleven roles covering
file loading, smell detection, planning, size assessment, LLM-based
transformation, quality verification, optional application, compilation
checking, and report generation. The workflow loads the target file, analyzes
static-analysis evidence, prepares a smell-oriented refactoring plan, checks
size feasibility, invokes the LLM refactoring agent, and forwards the generated
candidate to verification.

Planning combines deterministic rule-based guidance with optional LLM-based
refinement. The rule-based planner maps prioritized code smells to
refactoring-oriented instructions, while the LLM refactoring agent receives the
characterized task package and generates a candidate transformation. LLM
configurations are accessed through the same workflow interface, keeping the
orchestration fixed while allowing model-specific generation behaviour to vary.
If the candidate is unchanged, REFINE can retry with additional feedback;
verification failures are recorded as evidence rather than triggering a full
re-planning loop.

\subsection{Verification Trace}
\label{sec:refine-verification}

After candidate generation, REFINE re-analyzes the transformed file before
retaining it as a refactoring candidate. Let $B$ and $A$ denote the number of
detected code smells before and after refactoring, respectively. REFINE records
absolute code-smell reduction as:
\begin{equation}
\Delta_{\text{smell}} = B - A.
\end{equation}
A positive value indicates that the candidate reduces detected code smells,
whereas a negative value indicates that additional code smells were introduced.
REFINE also computes relative code-smell improvement:
\begin{equation}
\text{Improvement}_{\text{smell}} =
\begin{cases}
0, & \text{if } B = 0,\\[1mm]
\dfrac{B-A}{B} \times 100, & \text{otherwise.}
\end{cases}
\end{equation}

REFINE then applies static verification gates before either marking a candidate
as passing the configured checks in interactive configurations or retaining it
as evidence in the artifacts-only study configuration. A candidate is checked
for source difference, code-smell non-worsening, public API preservation,
exception-handling risks such as empty or comment-only catch blocks, configured
size bounds, and baseline method preservation. Source difference is based on
whitespace-normalized similarity, size preservation uses line-ratio bounds, and
method preservation requires retaining at least 85\% of baseline methods. These
checks are static preservation proxies and should not be interpreted as proof of
behavioral equivalence.

For each model pass, REFINE persists the original and generated source,
agent-step traces, before--after metrics, model metadata, verification
diagnostics, rejection reasons, and saved reports. This trace links each
refactoring decision to the evidence produced during characterization,
orchestration, and verification, and supports later analysis of code-smell
movement, quality indicators, change footprint, and preservation risks.
\section{Experimental Design}
\label{sec:experimental-design}

\subsection{Research Questions}
\label{sec:research-questions}

We study \textsc{REFINE} through three Research Questions (RQs) that evaluate
observed file-level outcomes, preservation evidence, and refactoring behaviour.
The matched direct-prompt baseline is used as comparative context rather than
as a separate research question.

\begin{itemize}
  \item \textbf{RQ1:}
  What before--after changes in detected code smells and automated quality
  indicators are observed after applying the multi-agent workflow?
  
  \item \textbf{RQ2:}
  What behavioural-risk and structural-preservation indicators are retained
  after refactoring, and what residual preservation risks remain?
  
  \item \textbf{RQ3:}
  What refactoring behaviours are associated with code-smell reduction across
  LLM configurations?
\end{itemize}

RQ1 examines target-level code-smell reduction and movement in automated
quality indicators. RQ2 examines static behavioural-risk proxies and failure
diagnostics. RQ3 examines how reductions are achieved through behaviours such
as code addition, code removal, method extraction, method renaming, class
splitting, duplicate-code removal, and public-method changes.

\subsection{Study Scope and Comparative Context}
\label{sec:study-scope}

This study evaluates \textsc{REFINE} as a controlled Java file-level workflow
for generating and analysing refactoring candidates. The main study uses 450
Java files from 15 systems and applies the same orchestration, prompting, and
verification rules across three LLM configurations. The purpose is to examine
what the workflow produces under controlled file-level conditions, including
code-smell reduction, quality movement, preservation evidence, and refactoring
behaviour.

To provide comparative context, we include a direct-prompt baseline on a
matched 150-file subset. This baseline compares \textsc{REFINE} with a simpler
strategy in which the same LLM configurations are directly prompted to refactor
the same original files. The baseline supports claims about this matched subset
only and is not intended as a comprehensive benchmark against all refactoring
tools, prompt designs, or agentic architectures.

The study is intentionally file-level. Each selected Java file is treated as an
independent refactoring unit, and the LLM receives the target file, prioritized
code smells, refactoring instructions, public API constraints, and retry notes
when applicable. Cross-file dependencies, tests, build configurations, and
repository-wide context are not included in the LLM prompt.

\subsection{Subject Systems and File Selection}
\label{sec:subject-systems}

We evaluate \textsc{REFINE} on 15 open-source Java systems selected to provide
variation in project size, application domain, and codebase structure. The
systems were required to be primarily implemented in Java, publicly available,
non-trivial, suitable for static analysis, and large enough to support
file-level sampling. Table~\ref{tab:dataset-overview} summarizes the selected
subject systems.

\begin{table}[t]
\centering
\caption{Dataset overview}
\label{tab:dataset-overview}
\tiny
\setlength{\tabcolsep}{2.5pt}
\renewcommand{\arraystretch}{1.05}
\resizebox{\columnwidth}{!}{%
\begin{tabular}{|r|l|c|c|c|c|}
\hline
\rowcolor{gray!15}
\textbf{\#} & \textbf{Project name} & \textbf{\# Java files} & \textbf{\# Classes} & \textbf{\# Methods} & \textbf{KLOC} \\
\hline
1  & JHotDraw            & 685   & 773   & 5{,}897  & 122 \\
2  & GanttProject        & 570   & 837   & 4{,}727  & 86 \\
3  & ArgoUML             & 1{,}687 & 2{,}017 & 11{,}846 & 332 \\
4  & Apache Xerces-J     & 1{,}109 & 1{,}308 & 8{,}004  & 297 \\
5  & Apache Ant          & 993   & 1{,}365 & 10{,}184 & 239 \\
6  & JEdit               & 615   & 1{,}143 & 5{,}883  & 193 \\
7  & JUnit 4             & 219   & 246   & 1{,}179  & 20 \\
8  & SLF4J               & 155   & 151   & 912      & 20 \\
9  & Logback             & 661   & 660   & 2{,}735  & 59 \\
10 & Mockito             & 521   & 587   & 2{,}138  & 48 \\
11 & Guava               & 1{,}995 & 4{,}139 & 21{,}018 & 475 \\
12 & jsoup               & 97    & 206   & 1{,}250  & 29 \\
13 & Eclipse Collections & 997   & 1{,}360 & 16{,}502 & 214 \\
14 & Checkstyle          & 2{,}273 & 2{,}839 & 7{,}215  & 188 \\
15 & JabRef              & 2{,}707 & 3{,}573 & 20{,}262 & 516 \\
\hline
\end{tabular}%
}
\end{table}

We consider production Java source files only. Test paths, generated build
outputs, version-control folders, and \textsc{REFINE}-internal folders are
excluded from the eligible file pool. KLOC denotes physical lines of production
Java source code, rounded to thousands; class and method counts are lightweight
static scale indicators.

From the eligible files, we selected refactoring units using stratified random
sampling over a code-smell-count $\times$ LOC grid. A file was eligible if it
contained at least one baseline code smell from the configured backend. The
smell-count strata distinguish lightly, moderately, and heavily affected files,
while the LOC dimension prevents domination by a single size range. The default
sampling seed was 42, and each batch selected at most 15 files while avoiding
previously selected paths. This process yielded 450 Java files; because each
file was refactored independently by three LLM configurations, the
complete-case cohort contains 1,350 model-pass outputs.

\subsection{Contextual Direct-Prompt Baseline}
\label{sec:baseline-design}

To assess whether \textsc{REFINE} provides value beyond directly prompting the
same LLMs, we evaluate a direct-prompt baseline on a matched 150-file subset.
The subset contains 10 files from each of the 15 subject systems. Each selected
file is evaluated with the same three LLM configurations, producing 450 matched
file-model pairs.

For each pair, \textsc{REFINE} and direct prompting start from the same
original Java file and the same LLM configuration. The direct-prompt baseline
receives the target file and an instruction to reduce the detected code smells
while preserving behaviour-relevant structures and public APIs. Unlike
\textsc{REFINE}, it does not use the multi-agent planning, orchestration,
verification-gate, or retry workflow. After generation, both outputs are
evaluated using the same post-hoc analysis pipeline.

\subsection{Experimental Setting}
\label{sec:experimental-setting}

The experiment was executed using the \textsc{REFINE} implementation described
in Section~\ref{sec:approach}. In this empirical study, code smells were detected and re-counted before and
after refactoring using PMD~7.10.0~\cite{pmd710} on Java~17 with the
\texttt{refactai-pmd-ruleset.xml} ruleset. Therefore, the
reported reductions should be interpreted as reductions in detected code smells
under this configuration, not as manually validated Fowler-style smell
instances.

We evaluated three LLM configurations through OpenRouter, using
\texttt{openai/gpt-5.5}, \texttt{google/gemini-3.1-pro-preview}, and
\texttt{anthropic/claude-opus-4.8}. Each configuration received the same
file-level task, including the complete target Java file, prioritized code
smells, rule-based refactoring instructions, public API constraints, and retry
notes when applicable.


All runs used the \texttt{independent\_parallel} mode: for each selected file,
the three LLM configurations were executed independently from the same frozen
baseline using the same \textsc{REFINE} prompt template, workflow settings, and
verification procedure. Before LLM invocation, \textsc{REFINE} applied a
160{,}000 estimated input-token preflight budget and line-count feasibility
checks. The study used an artifacts-only protocol in which generated candidates,
model metadata, before--after measurements, and verification traces were
persisted for replication, but candidate files were not written back to the live
workspace.
\subsection{Measures}
\label{sec:measures}

We measure each model pass using four groups of automatically collected
indicators. For \textbf{RQ1}, we use before--after counts of detected code
smells, severity-level changes, rule-level code-smell resolution, and static
quality proxies such as LOC, complexity, maintainability, testability, Halstead
metrics, and cohesion. Code-smell reduction and relative improvement follow the
definitions in Section~\ref{sec:refine-verification}.

For \textbf{RQ2}, we use verification outcomes, rejection reasons, and
preservation-risk proxies, including public API preservation,
method-signature preservation, exception-handling preservation,
framework-contract preservation, conditional-logic preservation, critical
assert/fail-call preservation, and a complete static preservation proxy. These
indicators are static proxies and do not prove behavioral equivalence.

For \textbf{RQ3}, we measure change footprint and refactoring behaviour using
lines added, lines removed, churn, method extraction, method renaming, class
splitting, duplicate-removal proxies, public API changes, and addition- or
deletion-heavy edit patterns. For the direct-prompt baseline, we use the same
outcome families where applicable: analyzability, code-smell reduction,
preservation indicators, public-method removal, and change-footprint metrics.

\subsection{Data Analysis}
\label{sec:data-analysis}

We analyze the main study at the model-pass level and report pooled and
model-wise descriptive statistics. For RQ1 before--after comparisons, we use
Wilcoxon signed-rank tests with Holm adjustment and report rank-biserial effect
sizes~\cite{wilcoxon1945,holm1979,cureton1956}. For matched binary
preservation indicators in RQ2, we use Cochran's $Q$ test for model-level
differences and Holm-adjusted McNemar tests for pairwise comparisons; Wilson
score 95\% confidence intervals are reported for pass-rate
estimates~\cite{cochran1950,mcnemar1947,wilson1927}. For RQ3 behaviour
comparisons over the code-smell-reducing subset, we use Kruskal--Wallis tests
because the subset is not fully matched after filtering to outputs that reduced
detected code smells. Spearman rank correlations are used to examine
associations between refactoring behaviours and absolute and relative
code-smell reduction~\cite{kruskal1952,spearman1904}.

For the contextual direct-prompt baseline, numeric paired outcomes are compared
using Wilcoxon signed-rank tests, and binary paired outcomes are compared using
McNemar tests. The baseline is reported separately from the main RQ1--RQ3
analysis because it uses the 150-file matched subset rather than the full
450-file cohort.

Failed or incomplete passes are reported separately from verifier rejections.
Because files are nested within systems, we interpret the inferential tests as
file-level evidence and do not make project-level causal claims. System-level
profiles are reported in the results to show whether aggregate patterns are
broadly distributed across subject systems or concentrated in a few projects.

\section{Results}
\label{sec:results}
This section reports the empirical results for the three research questions and
the contextual direct-prompt baseline, using the dataset and supporting
artifacts available online~\cite{REFINEDataset}.
\subsection{Observed File-Level Outcomes (RQ1)}
\label{sec:rq1-results}

\begin{figure*}[t]
    \centering
    \includegraphics[width=0.75\linewidth]{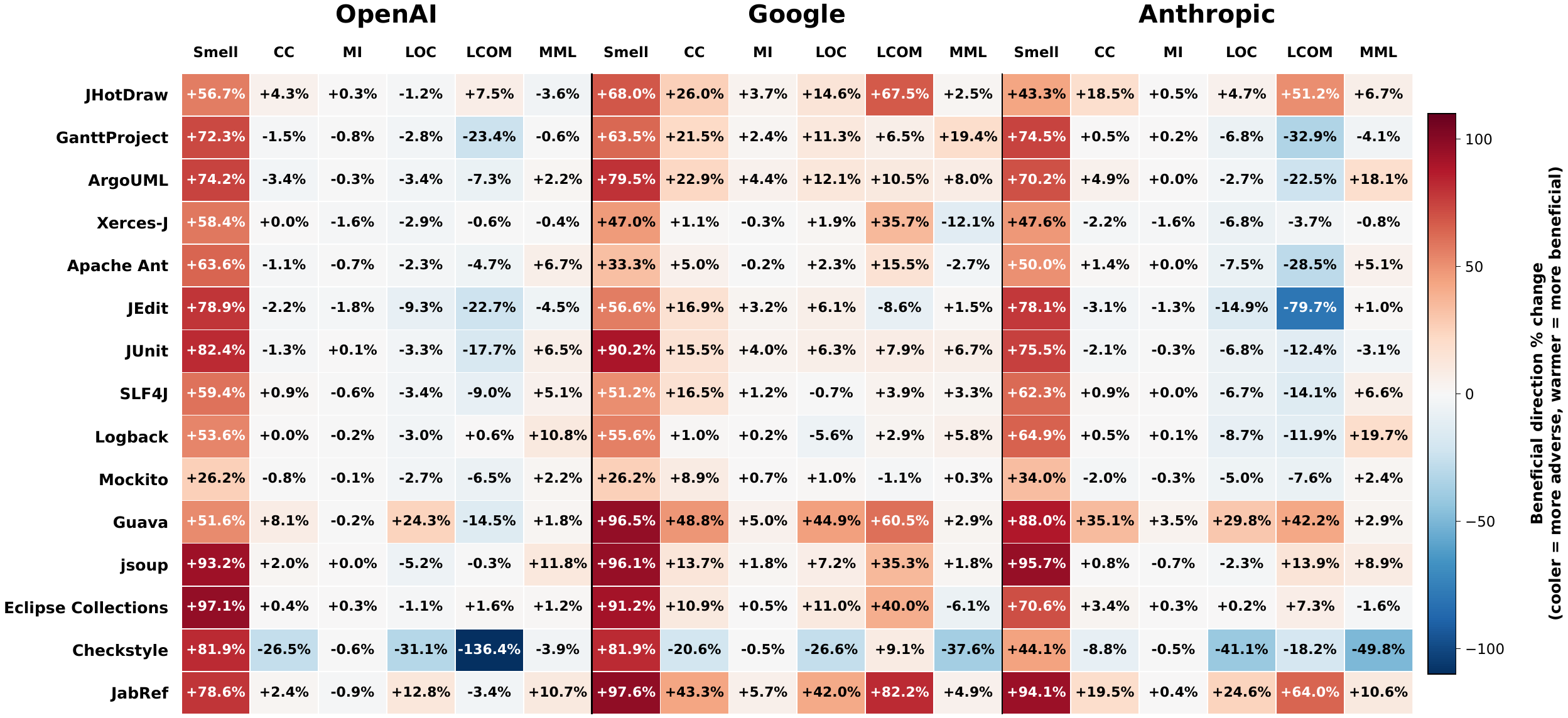}
    \caption{System-level profile for RQ1. Cells show beneficial-direction percentage change for each system, model, and indicator. Positive values are beneficial; negative values are adverse. Total Code Smells, CC, LOC, LCOM, and MML are lower-is-better; MI is higher-is-better. Warmer colours indicate stronger beneficial movement.}
    \label{fig:rq1-system-heatmap}
\end{figure*}

To answer RQ1, we compare before--after measurements across the three LLM
configurations. Table~\ref{tab:rq1-summary} reports aggregate changes in
detected code smells, severity levels, automated quality indicators, and
structural metrics. Figure~\ref{fig:rq1-system-heatmap} shows whether these
changes are distributed across systems or concentrated in a few cases.

\faIcon{thumbs-up}~\textbf{Code-smell reduction.}
REFINE reduces detected code smells across all three LLM configurations.
Total code-smell reduction ranges from 68.26\% to 72.79\%, with statistically
significant reductions for all severity categories. The largest reductions are
observed for major smells, ranging from 86.51\% to 91.60\%. This indicates that
the workflow is most effective when the target is explicit and measurable
through static smell evidence.

\begin{table*}[t]
\centering
\caption{Aggregate before--after results for code-smell reduction, quality indicators, and structural metrics.}
\label{tab:rq1-summary}
\scriptsize
\setlength{\tabcolsep}{2.0pt}
\renewcommand{\arraystretch}{0.92}
\resizebox{\textwidth}{!}{%
\begin{tabular}{
llc r
!{\vrule width 0.6pt}
rrr
!{\vrule width 0.6pt}
rrr
!{\vrule width 0.6pt}
rrr
!{\vrule width 0.6pt}
rrr
}
\toprule
\textbf{Group} & \textbf{Metric} & \textbf{Dir.} &
\textbf{$\Sigma$ Before} &
\multicolumn{3}{c}{\textbf{$\Sigma$ After}} &
\multicolumn{3}{c}{\textbf{$\Sigma\Delta=\Sigma$Before$-\Sigma$After}} &
\multicolumn{3}{c}{\textbf{\% $\Delta$}} &
\multicolumn{3}{c}{\textbf{Holm $p$ / $r$}} \\
\cmidrule(lr){5-7}
\cmidrule(lr){8-10}
\cmidrule(lr){11-13}
\cmidrule(lr){14-16}
& & & &
\textbf{GPT-5.5} & \textbf{Gemini 3.1} & \textbf{Opus 4.8} &
\textbf{GPT-5.5} & \textbf{Gemini 3.1} & \textbf{Opus 4.8} &
\textbf{GPT-5.5} & \textbf{Gemini 3.1} & \textbf{Opus 4.8} &
\textbf{GPT-5.5} & \textbf{Gemini 3.1} & \textbf{Opus 4.8} \\
\midrule

\multirow{4}{*}{\textbf{SMELLS}}
& Total Code Smells    & $\downarrow$ & 3\,021 & 959 & 822 & 952 & 2\,062 & 2\,199 & 2\,069 & \textbf{68.26} & \textbf{72.79} & \textbf{68.49} & $\mathbf{p<.001,\ r=.986}$ & $\mathbf{p<.001,\ r=1.000}$ & $\mathbf{p<.001,\ r=.992}$ \\
& Critical Code Smells & $\downarrow$ & 130 & 42 & 28 & 64 & 88 & 102 & 66 & 67.69 & 78.46 & 50.77 & $p<.001,\ r=.600$ & $p<.001,\ r=.627$ & $p<.001,\ r=.602$ \\
& Major Code Smells    & $\downarrow$ & 2\,106 & 272 & 177 & 284 & 1\,834 & 1\,929 & 1\,822 & \textbf{87.08} & \textbf{91.60} & \textbf{86.51} & $\mathbf{p<.001,\ r=.979}$ & $\mathbf{p<.001,\ r=.998}$ & $\mathbf{p<.001,\ r=.990}$ \\
& Minor Code Smells    & $\downarrow$ & 785 & 239 & 84 & 213 & 546 & 701 & 572 & 69.55 & 89.30 & 72.87 & $p<.001,\ r=.962$ & $p<.001,\ r=.992$ & $p<.001,\ r=.966$ \\

\midrule

\multirow{4}{*}{\textbf{QUALITY}}
& Cyclomatic Complexity & $\downarrow$ & 4\,631 & 4\,606 & 3\,654 & 4\,301 & 25 & 977 & 330 & 0.54 & \textbf{21.10} & 7.13 & $p=1.000,\ r=-.010$ & $\mathbf{p<.001,\ r=.709}$ & $p=1.000,\ r=.128$ \\
& Maintainability Index & $\uparrow$ & 34\,019.9 & 33\,878.3 & 34\,702.9 & 34\,032.7 & 141.6 & -683.0 & -12.8 & 0.42 & -2.01 & -0.04 & $p=1.000,\ r=.361$ & $p=1.000,\ r=-.060$ & $p=1.000,\ r=.091$ \\
& Testability Index & $\uparrow$ & 20\,414.6 & 19\,197.4 & 19\,801.0 & 19\,944.3 & 1\,217.2 & 613.6 & 470.3 & 5.96 & 3.01 & 2.30 & $p=1.000,\ r=.600$ & $p=1.000,\ r=.197$ & $p=1.000,\ r=.369$ \\
& Halstead Effort & $\downarrow$ & 281\,985\,379.7 & 278\,953\,620.4 & 159\,234\,977.7 & 201\,221\,901.9 & 3\,031\,759.3 & 122\,750\,402.0 & 80\,763\,477.8 & 1.08 & 43.53 & 28.64 & $p=1.000,\ r=-.476$ & $p=1.000,\ r=-.026$ & $p=1.000,\ r=-.449$ \\

\midrule

\multirow{3}{*}{\textbf{STRUCTURE}}
& Lines of Code & $\downarrow$ & 91\,574 & 89\,673 & 77\,151 & 88\,562 & 1\,901 & 14\,423 & 3\,012 & 2.08 & 15.75 & 3.29 & $p=1.000,\ r=-.758$ & $p=1.000,\ r=-.514$ & $p=1.000,\ r=-.852$ \\
& LCOM & $\downarrow$ & 30\,941 & 31\,983 & 17\,930 & 25\,305 & -1\,042 & 13\,011 & 5\,636 & -3.37 & \textbf{42.05} & 18.22 & $p=1.000,\ r=-.605$ & $\mathbf{p=.002,\ r=.392}$ & $p=1.000,\ r=-.575$ \\
& Mean Method Length & $\downarrow$ & 7\,614.7 & 7\,501.5 & 6\,858.9 & 7\,503.6 & 113.2 & 755.8 & 111.1 & 1.49 & 9.93 & 1.46 & $p=1.000,\ r=.001$ & $p=1.000,\ r=-.342$ & $p=1.000,\ r=-.022$ \\

\bottomrule
\end{tabular}%
}

\vspace{0.15em}

{\tiny\textit{Note.} $\Sigma$ values aggregate all file-level observations.
$\%\Delta=(\Sigma\Delta/\Sigma\text{Before})\times100$. Positive changes are
beneficial for $\downarrow$ metrics; negative changes are beneficial for
$\uparrow$ metrics. $p$ denotes Holm-adjusted Wilcoxon results and $r$ the
rank-biserial effect size. Bold values mark statistically significant results
discussed in the text.}

\end{table*}

\faIcon{chart-line}~\textbf{Quality and structural indicators.}
Quality and structural indicators show mixed movement. Gemini 3.1 shows the
clearest secondary gains, with significant reductions in cyclomatic complexity
and LCOM. Other indicators, including maintainability, testability, Halstead
effort, LOC, and mean method length, are not consistently improved across
models. This separates targeted smell reduction from broader quality
improvement: the workflow reduces the measured smell targets, but secondary
quality effects remain metric- and model-dependent.

\faIcon{th}~\textbf{System-level profile.}
Figure~\ref{fig:rq1-system-heatmap} shows that code-smell reduction appears
across all 15 systems. Non-smell indicators vary more by system and model.
Gemini 3.1 shows the broadest beneficial pattern, while GPT-5.5 and Opus 4.8
show less stable structural effects. Thus, the aggregate smell-reduction result
is not driven by a single system, but broader quality movement is less uniform.

\begin{rqanswer}
\noindent\faIcon{lightbulb}~\textbf{Answer to RQ1:}
After applying REFINE, detected code smells decrease substantially across all
three LLM configurations, with the largest reductions observed for major smells.
Automated quality indicators show mixed before--after changes: cyclomatic
complexity and LCOM improve mainly for Gemini 3.1, while maintainability,
testability, Halstead effort, LOC, and mean method length do not improve
consistently across models.
\end{rqanswer}

\subsection{Preservation and Verification (RQ2)}
\label{sec:rq2-results}

To answer RQ2, we analyse preservation indicators, verification outcomes, and
failure diagnostics. Table~\ref{tab:rq2-preservation} reports pass rates for
static preservation checks, while Table~\ref{tab:rq2-failure-diagnostics}
summarizes the main failure signals.

\faIcon{shield-alt}~\textbf{Preservation indicators.}
Most static preservation indicators pass at high rates. Public method
signatures, exception-handling constructs, framework contracts, and conditional
constructs are preserved in most outputs. The complete static preservation proxy
ranges from 81.8\% to 90.0\%. Gemini 3.1 is lower than the other models on
several indicators. These results indicate that many candidates retain visible
API- and structure-related elements, but the checks remain static proxies and
do not prove behavioural equivalence.

\faIcon{exclamation-triangle}~\textbf{Residual risks.}
The weakest indicator is critical assert/fail-call preservation, which passes
in only 57.1\% of outputs for each model. This shows a common preservation risk
across all evaluated LLM configurations rather than a model-specific issue.
Table~\ref{tab:rq2-failure-diagnostics} also shows that critical assert/fail-call
failures account for 579 model-pass outputs. Therefore, smell reduction must be
interpreted together with preservation evidence.

\begin{table*}[t]
\centering
\caption{Automated preservation and verification pass rates across LLM configurations.}
\label{tab:rq2-preservation}
\scriptsize
\setlength{\tabcolsep}{2.2pt}
\renewcommand{\arraystretch}{0.92}
\resizebox{\textwidth}{!}{%
\begin{tabular}{
l
!{\vrule width 0.6pt}
ccc
!{\vrule width 0.6pt}
ccrll
}
\toprule
\textbf{Indicator} &
\multicolumn{3}{c}{\textbf{LLM configuration result, $n/450$ (\%)}} &
\multicolumn{5}{c}{\textbf{Model comparison and statistical evidence}} \\
\cmidrule(lr){2-4}
\cmidrule(lr){5-9}
&
\textbf{GPT-5.5} & \textbf{Gemini 3.1} & \textbf{Opus 4.8} &
\textbf{Pooled Wilson 95\% CI} & \textbf{Cochran $Q$} & \textbf{$p$} &
\textbf{Best model} & \textbf{Pairwise (Holm--McNemar)} \\
\midrule

Public method signatures preserved
& \textbf{420/450 (93.3)} & 388/450 (86.2) & \textbf{424/450 (94.2)}
& [89.6, 92.7] & 44.08 & $<.001$
& Opus 4.8 & Gemini 3.1 lower than GPT-5.5 and Opus 4.8 \\

Exception-handling constructs preserved
& \textbf{431/450 (95.8)} & 416/450 (92.4) & \textbf{433/450 (96.2)}
& [93.5, 95.9] & 15.70 & $<.001$
& Opus 4.8 & Gemini 3.1 lower than GPT-5.5 and Opus 4.8 \\

Framework contracts preserved
& \textbf{425/450 (94.4)} & 407/450 (90.4) & 420/450 (93.3)
& [91.2, 94.0] & 21.58 & $<.001$
& GPT-5.5 & Gemini 3.1 lower than GPT-5.5 and Opus 4.8 \\

Conditional constructs preserved
& \textbf{405/450 (90.0)} & 387/450 (86.0) & 395/450 (87.8)
& [86.1, 89.6] & 9.57 & .008
& GPT-5.5 & Gemini 3.1 lower than GPT-5.5 \\

Critical assert/fail-call preservation
& \textbf{257/450 (57.1)} & \textbf{257/450 (57.1)} & \textbf{257/450 (57.1)}
& [54.5, 59.7] & 0.00 & 1.000
& -- & No model difference \\

Complete static preservation proxy
& \textbf{396/450 (88.0)} & \textbf{368/450 (81.8)} & \textbf{405/450 (90.0)}
& [84.7, 88.3] & 37.86 & $<.001$
& Opus 4.8 & Gemini 3.1 lower than GPT-5.5 and Opus 4.8 \\

Refactoring success
& \textbf{439/450 (97.6)} & 416/450 (92.4) & \textbf{428/450 (95.1)}
& [93.7, 96.1] & 20.36 & $<.001$
& GPT-5.5 & Gemini 3.1 lower than GPT-5.5 \\

\bottomrule
\end{tabular}%
}

\vspace{0.15em}

{\scriptsize\textit{Note.} Pass rates are computed over 450 outputs per model.
Wilson 95\% CIs are pooled across models. Cochran's $Q$ tests matched
model-level differences; pairwise comparisons use Holm-adjusted McNemar tests.
These indicators are static preservation proxies, not behavioural-equivalence
proofs.}
\end{table*}

\faIcon{bug}~\textbf{Failure diagnostics.}
The clearest concrete diagnostic signal is public-method removal. All failures
with concrete diagnostic text refer to this issue. Gemini 3.1 has the highest
number of public-method removal cases and the highest mean number of removed
public methods per affected file. Critical assert/fail-call loss is the largest
failure category, but it is less explained by the available diagnostics. This
means public API deletion is the most concrete observed failure mechanism,
whereas assert/fail-call loss remains the largest unresolved risk signal.

\begin{table}[t]
\centering
\caption{Failure diagnostics for behavioural-risk preservation indicators.}
\label{tab:rq2-failure-diagnostics}
\scriptsize
\setlength{\tabcolsep}{2.6pt}
\renewcommand{\arraystretch}{0.95}
\resizebox{\columnwidth}{!}{%
\begin{tabular}{lcc}
\toprule
\textbf{Diagnostic} & \textbf{Overall} & \textbf{Model signal} \\
\midrule

\multicolumn{3}{l}{\textbf{Per-indicator failures}} \\
\midrule

Critical assert/fail-call failures
& \textbf{579 (42.9\%)} & \textbf{193 per model} \\

Conditional-construct failures
& 163 (12.1\%) & Gemini 3.1 highest: 63/450 \\

Public-method signature failures
& 118 (8.7\%) & Gemini 3.1 highest: 62/450 \\

Framework-contract failures
& 98 (7.3\%) & Gemini 3.1 highest: 43/450 \\

Exception-handling failures
& 70 (5.2\%) & Gemini 3.1 highest: 34/450 \\

\midrule

\multicolumn{3}{l}{\textbf{Concrete diagnostic signal}} \\
\midrule

Failures with concrete diagnostic text
& \textbf{75} & all public-method removals \\

Model-level public-method removal cases
& \textbf{146} & GPT-5.5: 41; \textbf{Gemini 3.1: 71}; Opus 4.8: 34 \\

Mean public methods removed per affected file
& -- & GPT-5.5: 0.40; \textbf{Gemini 3.1: 1.45}; Opus 4.8: 0.60 \\

\bottomrule
\end{tabular}%
}

\vspace{0.15em}

{\scriptsize\textit{Note.} Percentages use 1,350 model-pass outputs as denominator.
Concrete diagnostic text is available only for a subset of failures.}

\end{table}

\begin{rqanswer}
\noindent\faIcon{lightbulb}~\textbf{Answer to RQ2:}
After refactoring, most behavioural-risk and structural-preservation indicators
are retained at high rates, including public method signatures,
exception-handling constructs, framework contracts, conditional constructs, and
the complete static preservation proxy. The main residual risks are weak
critical assert/fail-call preservation and public-method removal, with
public-method removal appearing most clearly in Gemini 3.1 outputs.
\end{rqanswer}
\subsection{Refactoring Behaviour (RQ3)}
\label{sec:rq3}

To answer RQ3, we analyse the 990 model-pass outputs that reduced detected code
smells. Table~\ref{tab:rq3-refactoring-behaviour} reports model-level
refactoring behaviour and correlations with absolute and relative code-smell
reduction.

\faIcon{exchange-alt}~\textbf{Model-specific edit profiles.}
The models reduce smells through different edit profiles. GPT-5.5 produces the
most compact edits. Gemini 3.1 is deletion-heavy, with the largest line and
method removal. Opus 4.8 is extraction-heavy, with the largest number of method
extractions and the highest churn rate. These profiles show that similar
smell-reduction outcomes can be produced through different transformation
styles.

\faIcon{chart-line}~\textbf{Behaviour--reduction associations.}
Edit volume is positively associated with absolute smell reduction. Total churn,
lines added, and lines removed correlate with the number of smells removed.
However, these behaviours are negatively associated with relative smell
reduction. Larger edits therefore remove more smells in raw counts, but not
necessarily a larger proportion of smells. This pattern suggests that high-churn
outputs often occur on larger or more smell-dense files.

\faIcon{code-branch}~\textbf{Less explanatory behaviours.}
Duplicate-code removal, method addition, and class splitting show limited
association with smell reduction. Method renaming differs in total counts, but
the per-pass model difference is not statistically significant. Thus, the main
observed behaviours associated with smell reduction are edit volume, removal,
and extraction rather than all recorded refactoring actions.

\begin{table}[t]
\centering
\caption{Refactoring behaviour and code-smell reduction associations.}
\label{tab:rq3-refactoring-behaviour}
\scriptsize
\setlength{\tabcolsep}{2.2pt}
\renewcommand{\arraystretch}{1.03}
\resizebox{\columnwidth}{!}{%
\begin{tabular}{lrrrrrr}
\toprule
&
\multicolumn{5}{c}{\textbf{Model-level refactoring behaviour}} &
\multicolumn{1}{c}{\textbf{Rank assoc.}} \\
\cmidrule(lr){2-6}
\cmidrule(lr){7-7}
\textbf{Refactoring behaviour} &
\textbf{GPT-5.5} &
\textbf{Gemini 3.1} &
\textbf{Opus 4.8} &
\textbf{KW $H$} &
\textbf{$p$} &
\textbf{$\rho_R/\rho_\%$} \\
\midrule

\multicolumn{7}{l}{\textit{Churn metrics (per-file means)}} \\
Total churn (lines) & \textbf{47.0} & 69.0 & \textbf{75.2} & 23.929 & $<.001$ & \textbf{+0.695 / -0.278} \\
Lines added & 21.2 & 14.3 & \textbf{31.6} & 46.969 & $<.001$ & \textbf{+0.646 / -0.351} \\
Lines removed & 25.8 & \textbf{54.6} & 43.6 & 12.107 & .002 & \textbf{+0.623 / -0.194} \\
Net LOC change & \textbf{-4.7} & \textbf{-40.3} & -11.9 & 55.745 & $<.001$ & +0.327 / -0.343 \\
Churn rate (\%) & 27.68 & 26.46 & \textbf{36.26} & 32.873 & $<.001$ & +0.230 / -0.063 \\
\midrule

\multicolumn{7}{l}{\textit{Structural actions (totals across code-smell-reducing outputs per model)}} \\
Methods extracted & 243 & 167 & \textbf{610} & \textbf{89.692} & $\mathbf{<.001}$ & +0.188 / -0.210 \\
Methods renamed & 132 & 541 & 288 & 1.559 & .459 & +0.327 / +0.051 \\
Class splits & 84 & 55 & 91 & 3.226 & .199 & +0.028 / -0.144 \\
Duplicate-code removals & 0 & 0 & 0 & -- & -- & +0.000 / +0.000 \\
\midrule

\multicolumn{7}{l}{\textit{Public API surface and static preservation}} \\
Methods added & 49 & 44 & 45 & 1.077 & .584 & +0.015 / +0.031 \\
Methods removed & 142 & \textbf{562} & 255 & 11.897 & .003 & +0.265 / +0.158 \\
Static preservation score (\%) & 93.95 & \textbf{89.87} & \textbf{94.29} & 8.606 & .014 & -0.235 / -0.069 \\
\bottomrule
\end{tabular}%
}

\vspace{1mm}
\begin{flushleft}
\scriptsize
\textit{Note.} Results use 990 outputs that reduced detected smells. Churn
metrics are per-file means; structural/API actions are model totals. KW denotes
Kruskal--Wallis tests; $\rho_R$ and $\rho_\%$ are Spearman correlations with
absolute and relative smell reduction.
\end{flushleft}
\end{table}

\begin{rqanswer}
\noindent\faIcon{lightbulb}~\textbf{Answer to RQ3:}
Code-smell reduction is associated mainly with edit volume, code removal,
method removal, and method extraction. Across models, GPT-5.5 reduces smells
with more compact edits, Gemini 3.1 with deletion-heavy edits, and Opus 4.8
with extraction-heavy edits. Larger edits remove more smells in absolute terms,
but not as a larger proportion of the original smell count.
\end{rqanswer}

\subsection{Contextual Direct-Prompt Baseline}
\label{sec:baseline-results}
To assess whether \textsc{REFINE} adds value beyond direct prompting, we compare
it with a direct-prompt baseline on the matched 150-file subset, yielding 450
file-model pairs. Table~\ref{tab:baseline-comparison} shows that \textsc{REFINE}
achieves higher median total smell reduction than direct prompting
(100.0\% vs. 20.8\%, $p<.001$). It also preserves public signatures more often,
removes fewer public methods, and produces smaller edits.

Direct prompting preserves critical assert/fail constructs more often than
\textsc{REFINE}. Thus, \textsc{REFINE} improves controlled smell reduction and
edit footprint, but it does not dominate direct prompting on all preservation
indicators. This comparison supports the value of the workflow structure for
more controlled candidate generation, while also showing that stronger smell
reduction can still leave preservation risks.

\begin{table}[t]
\centering
\caption{Contextual comparison between REFINE and direct prompting.}
\label{tab:baseline-comparison}
\scriptsize
\setlength{\tabcolsep}{2.5pt}
\renewcommand{\arraystretch}{0.94}
\begin{tabular}{p{0.48\columnwidth}rrrr}
\toprule
\textbf{Outcome} & \textbf{REF.} & \textbf{Dir.} & \textbf{$\Delta$} & \textbf{$p$} \\
\midrule
Valid/analyzable outputs (\%) & 100.0 & 100.0 & 0.0 & 1.000 \\
Total smell reduction, median (\%) & \textbf{100.0} & 20.8 & 79.2 & $<.001$ \\
Major smell reduction, median (\%) & 100.0 & 100.0 & 0.0 & $<.001$ \\
Critical smell reduction, median (\%) & 100.0 & 100.0 & 0.0 & .006 \\
Public signatures preserved (\%) & \textbf{92.0} & 75.1 & 16.9 & $<.001$ \\
Critical assert/fail preserved (\%) & 58.0 & \textbf{100.0} & -42.0 & $<.001$ \\
Complete preservation proxy (\%) & \textbf{87.8} & 0.0 & 87.8 & $<.001$ \\
Public-method removal cases & \textbf{46} & 112 & -66 & $<.001$ \\
Median total churn (LOC) & \textbf{14} & 65 & -51 & $<.001$ \\
Median net LOC change & 4 & 6 & -2 & .477 \\
Median lines added & \textbf{8} & 19 & -11 & $<.001$ \\
Median lines removed & \textbf{4} & 24 & -20 & $<.001$ \\
\bottomrule
\end{tabular}

\vspace{0.15em}
\parbox{\columnwidth}{%
\scriptsize\textit{Note.} REF. = REFINE; Dir. = direct prompting.
The comparison uses 450 matched file-model observations from the 150-file subset,
except major smell reduction ($n=396$) and critical smell reduction ($n=69$).
Binary outcomes use McNemar tests; numeric outcomes use Wilcoxon signed-rank tests.}
\end{table}
\section{Discussion}
\label{sec:discussion}


\subsection{Targeted Code-Smell Reduction Is Not General Quality Improvement}

The strongest observed outcome of \textsc{REFINE} is the reduction of detected
code smells. Across all three LLM configurations, the workflow reduces total
code-smell counts, with the largest reductions observed for major code smells.
This indicates that LLM-based refactoring can support measurable file-level
maintenance tasks when the target is explicit, operationalised through static
analysis, and re-checked after generation. This result is consistent with
recent evidence on LLM refactoring capability~\cite{cordeiro2024empirical} and
broader reviews of LLM-based code-quality improvement~\cite{alomari2025llms,martinez2025slr}.

At the same time, code-smell reduction should not be treated as a proxy for
general software quality improvement. Reductions in detected smells do not
transfer uniformly to maintainability, testability, Halstead effort, Lines of
Code, LCOM, or Mean Method Length. These indicators show mixed, adverse, or
non-significant changes depending on the model and system. This distinction
matters because code smells are useful maintainability indicators, but they do
not capture the full quality impact of refactoring~\cite{lacerda2020code}.
Evaluations of LLM-based refactoring should therefore report target-level smell
reduction together with secondary quality, structural, and preservation
evidence.

The matched direct-prompt baseline reinforces this point. \textsc{REFINE}
achieves stronger overall code-smell reduction with smaller edits and fewer
public-method removals, suggesting that workflow structure can guide LLMs
toward more controlled candidate generation. However, direct prompting preserves
critical assert/fail constructs more often. The comparison therefore supports a
multi-objective interpretation of LLM refactoring: smell reduction, edit
footprint, and preservation risk must be considered together.

\subsection{Preservation Evidence Exposes Residual Refactoring Risks}

The RQ2 results show that generated candidates retain several static
behavioural-risk indicators at high rates, including public method signatures,
exception-handling constructs, framework contracts, and conditional constructs.
However, preservation remains separate from code-smell mitigation. Critical
assert/fail-call preservation is weaker, and public-method removal is the
clearest concrete diagnostic signal. This distinction is central to
refactoring, which is traditionally understood as improving internal structure
while preserving externally observable behaviour~\cite{fowler2018refactoring,mens2004survey}.

The preservation evidence should therefore be interpreted as static risk
evidence, not proof of behavioural equivalence. A candidate may reduce detected
smells and pass several static checks while still changing behaviour through
removed assertions, altered control flow, changed public methods, or modified
framework interactions. The value of these checks is that they expose residual
risks for review. In practice, \textsc{REFINE}-style workflows should be treated
as candidate-generation and risk-filtering mechanisms, not fully autonomous
refactoring tools.

\subsection{Model Choice Shapes Refactoring Behaviour and Risk}

The RQ3 results show that the evaluated LLM configurations do not behave as
interchangeable refactoring engines. Although all three models reduce detected
code smells, they do so through different edit profiles. GPT-5.5 produces
comparatively compact changes, Gemini 3.1 shows a deletion-heavy profile with
stronger method removal and net LOC reduction, and Opus 4.8 shows an
extraction-heavy profile with substantially more method extractions. Thus, the
practical meaning of code-smell reduction depends not only on whether smells
are reduced, but also on how the reduction is achieved.

These model-specific behaviours reveal different trade-offs. Deletion-heavy
refactoring may reduce detected smells and simplify some metrics, but it can
increase API-surface risk when methods are removed. Extraction-heavy
refactoring may support decomposition, but it introduces more structural change
that requires review. Compact refactoring may be easier to inspect, but may
provide fewer secondary structural gains. This interpretation is consistent
with empirical evidence that LLM refactoring outcomes depend on model
behaviour, prompting, and validation context~\cite{cordeiro2024empirical,liu2025potential},
and with agentic-refactoring studies that emphasize evaluating what kinds of
changes agents produce~\cite{horikawa2025agentic,oueslati2025refagent}. Model
selection should therefore be treated as a refactoring-strategy decision, not
only a performance or cost choice.

\subsection{Evidence-Aware Workflows Provide Control but Not Autonomy}

The contextual baseline suggests that \textsc{REFINE} adds value over direct
prompting for the matched subset, especially in code-smell reduction,
public-signature preservation, public-method removal, and edit footprint.
However, this does not mean that multi-agent orchestration is universally
superior to all prompting strategies, single-agent workflows, or existing
refactoring tools. The baseline is contextual: it compares \textsc{REFINE} with
one direct-prompt strategy on the same files and LLM configurations.

These results also show that neither direct prompting nor the multi-agent
workflow should be interpreted as producing completed behaviour-preserving
refactorings. Behaviour preservation is central to the definition of
refactoring~\cite{fowler2018refactoring,mens2004survey}, yet recent empirical
studies show that LLM-based refactoring can still produce unsafe outputs,
including syntax errors or functionality-changing edits~\cite{liu2025potential}.
In our results, direct prompting preserves some critical assert/fail constructs
more often, whereas \textsc{REFINE} provides stronger code-smell reduction and
a more controlled edit footprint. However, both strategies leave preservation
risks. Thus, the main contribution of \textsc{REFINE} is not replacing
behaviour-preserving refactoring tools, but making LLM-generated refactoring
candidates more measurable, evidence-linked, and reviewable.

The broader value of \textsc{REFINE} lies in creating a controlled evaluation
object for LLM-based refactoring. Prior empirical work evaluates LLM
refactoring using combinations of code-smell reduction, static quality metrics,
and unit-test success~\cite{cordeiro2024empirical}, while recent repository-level
benchmarks emphasize compilation, test execution, and automated refactoring
detection as stronger validation evidence~\cite{xu2026swerefactor}. By separating
code-smell detection, planning, transformation, verification, and reporting,
\textsc{REFINE} links each generated candidate to its motivating evidence,
verification outcomes, and observed risks. This traceability enables more
detailed analysis than aggregate before--after improvement alone.

The boundary of this evidence is file-level. \textsc{REFINE} uses the target
file, detected code smells, refactoring instructions, public API constraints,
and verification traces, but it does not measure ripple effects across
dependent files, callers, tests, build configurations, or framework execution
paths. Therefore, moving from file-level candidates to repository-level
refactoring will require stronger project context, dependency analysis,
compilation, regression testing, and human-in-the-loop acceptance, consistent
with recent concerns that LLM refactoring remains error-prone and difficult to
fully automate without review~\cite{cordeiro2025llm}.
\subsection{Implications for Research and Practice}

For researchers, these results suggest that future evaluations of LLM-based
refactoring should move beyond aggregate improvement scores. Reporting only
code-smell reduction can hide differences in edit behaviour, preservation risk,
and structural side effects. Metrics such as churn, method extraction, method
removal, API-surface change, static preservation proxies, and direct-prompt
baselines help explain how an LLM achieves refactoring outcomes and what risks
come with those outcomes.

For practitioners, \textsc{REFINE} provides a traceable way to generate and
inspect Java file-level refactoring candidates, but it should not be treated as
an autonomous refactoring system. The results suggest that different models may
fit different maintenance goals: GPT-5.5 for compact edits, Gemini 3.1 for
aggressive size reduction with stronger API review, and Opus 4.8 for
decomposition-oriented refactoring. In all cases, generated candidates require
project-specific validation before integration.
\section{Related Work}
\label{sec:related-work}


\subsection{Automated and LLM-Based Refactoring}
\label{sec:automated-llm-refactoring}

Software refactoring is traditionally defined as improving the internal
structure of code while preserving externally observable behaviour~\cite{fowler2018refactoring,mens2004survey}.
Earlier research focused on tool-supported transformations, refactoring
opportunity identification, process automation, and quality metrics for
assessing maintainability improvements~\cite{mens2004survey,abid2020thirty}.
Code smells are widely used as indicators of maintainability problems, but
prior reviews emphasize that smell removal should not be treated as a complete
measure of software quality~\cite{lacerda2020code}. This distinction is
important for LLM-based refactoring because a generated change may reduce a
detected code smell while still introducing new quality or preservation risks.

Recent studies have explored LLMs for refactoring recommendation, code quality
improvement, and code transformation~\cite{alomari2025llms,martinez2025slr,
simoes2025refmodel}. Cordeiro et al.~\cite{cordeiro2024empirical} empirically
study LLM refactoring on Java projects and evaluate generated refactorings in
terms of code smells, static quality metrics, and unit-test success. Liu et
al.~\cite{liu2025potential} study general-purpose LLMs for identifying
refactoring opportunities and recommending refactoring solutions, showing that
LLMs can provide useful suggestions but may also produce unsafe outputs such as
syntax errors or functionality-changing edits. These results motivate
\textsc{REFINE}'s focus on generated refactoring candidates, before--after
quality measurement, and explicit preservation diagnostics rather than treating
LLM output as automatically safe.

\subsection{Agentic and Repository-Level Refactoring}
\label{sec:agentic-refactoring}

Agentic approaches aim to structure LLM-based refactoring through planning,
generation, validation, and repair. The closest related work is
RefAgent~\cite{oueslati2025refagent}, a multi-agent LLM-based framework for
automatic software refactoring that uses specialized agents for planning,
execution, testing, and iterative refinement. RefAgent reports improvements in
unit-test pass rate, smell reduction, and quality attributes, and compares its
multi-agent design with single-agent and non-agentic alternatives. This makes
RefAgent an important reference point for \textsc{REFINE}. In contrast,
\textsc{REFINE} does not aim to establish superiority over alternative
workflows; instead, it characterizes file-level outcomes, preservation risks,
and model-specific behaviours under one evidence-aware multi-agent workflow.

Recent repository-level and architectural studies show why this scope matters.
SWE-Refactor~\cite{xu2026swerefactor} introduces a repository-level benchmark
of developer-written Java refactorings validated through compilation, test
execution, and automated refactoring detection. It shows that realistic
refactoring evaluation requires repository context and that compound
refactorings remain difficult for current LLMs. SmellBench~\cite{dinu2026smellbench}
evaluates LLM agents on architectural code smell repair and shows that agents
still struggle with cross-module design understanding, repair aggressiveness,
and net codebase impact. These studies reinforce the boundary of our work:
\textsc{REFINE} studies targeted file-level code-smell mitigation, not safe
repository- or architecture-level refactoring.

\subsection{Conclusive Summary}
\label{sec:related-summary}

Overall, prior work shows that LLMs and agents can support refactoring, but
also that generated changes require evaluation beyond smell reduction alone.
\textsc{REFINE} complements this work by studying evidence-aware file-level
refactoring-candidate generation rather than treating LLM output as completed
behaviour-preserving refactoring.

Across 450 Java files, 15 systems, and 1,350 model-pass outputs,
\textsc{REFINE} connects code-smell reduction with quality indicators, static
preservation proxies, failure diagnostics, and model-specific refactoring
behaviour. Its contribution is not a general claim of superiority over direct
prompting, existing tools, or repository-level benchmarks, but a controlled
characterization of what an evidence-aware multi-agent workflow produces and
where preservation and integration risks remain.
\section{Threats to Validity}
\label{sec:threats}

\textbf{Construct validity:}
This study operationalises code smells using detections produced by the
configured static-analysis backend. In our empirical setting, this backend is
PMD; therefore, the reported reductions should be interpreted as reductions in
PMD-detected smells under the selected ruleset, not as complete removal of all
possible design smells or manually validated Fowler-style smell instances.
Similarly, the quality indicators used in RQ1, including maintainability,
testability, Halstead effort, LOC, LCOM, and mean method length, are automated
proxies and may not fully capture developer-perceived maintainability or design
quality. The preservation indicators used in RQ2 are also static proxies: they
capture changes to public APIs, exception handling, framework-related
constructs, conditional logic, and critical assert/fail calls, but they do not
prove behavioural equivalence. To reduce this threat, we report code-smell
reduction, quality indicators, preservation evidence, failure diagnostics, and
refactoring behaviour separately rather than treating any single metric as a
complete measure of refactoring effectiveness.

\textbf{Internal validity:}
The results may be influenced by implementation choices in \textsc{REFINE},
including the static-analysis ruleset, prompt template, retry policy,
verification thresholds, and evidence gates. Although all LLM configurations
were evaluated under the same orchestration and verification rules, outputs may
still vary because of model behaviour, service-side updates, or routing
infrastructure. We mitigated this threat by using a frozen baseline, identical
file-level task packages, fixed orchestration rules, independent model runs,
and persisted verification traces. Nevertheless, exact replication may be
affected by changes in the underlying LLM services. The study also used an
artifacts-only protocol, so the results reflect controlled candidate evaluation
rather than full project integration.

\textbf{External validity:}
The evaluation is limited to Java and file-level refactoring tasks. Although
the 15 subject systems cover different sizes and domains, they do not represent
all languages, frameworks, architectures, or industrial settings. The sample
contains production Java files with at least one baseline detected smell, so
the results may not generalise to clean files, test code, generated code, or
other maintenance tasks. The LLMs also receive file-level context rather than
full repository context. Therefore, the results should not be interpreted as
evidence of repository-wide or system-level refactoring effectiveness.

\textbf{Conclusion validity:}
The analysis focuses on model-pass and file-level outcomes. Files are nested
within systems and may not be fully independent, and aggregate improvements may
be influenced by larger or more smell-dense files. To reduce this threat, we
report model-wise results, system-level profiles, non-parametric tests,
multiple-comparison adjustments, and effect sizes where appropriate. Still,
statistical improvement in static indicators should not be interpreted as proof
of practical maintainability improvement, behavioural safety, or production
readiness.

\textbf{Contextual baseline scope:}
The matched direct-prompt baseline contextualizes whether \textsc{REFINE} adds
value beyond directly prompting the same LLMs on the same files. However, it is
limited to a 150-file subset and one direct-prompt strategy. It does not cover
all prompt designs, single-agent workflows, non-agentic pipelines, traditional
refactoring tools, or component ablations. Therefore, the baseline supports
claims about this matched subset only, not a general claim that \textsc{REFINE}
outperforms all alternative refactoring workflows.

\textbf{Behavioural safety and deployment readiness:}
The preservation checks are static diagnostics, not full behavioural
validation. Generated candidates were not evaluated through repository-level
compilation, regression testing, call-graph impact analysis, or human review as
part of the main experiment. Thus, even candidates that reduce detected smells
and pass configured checks may still introduce behavioural, integration, or
design-level risks. The results should be interpreted as evidence about
candidate generation and risk filtering, not autonomous production-ready
refactoring.
\section{Conclusion}
\label{sec:conclusion}

This paper presented \textsc{REFINE}, an evidence-aware multi-agent workflow for generating and evaluating Java file-level refactoring candidates. We evaluated \textsc{REFINE} on 450 Java files from 15 open-source systems, producing 1,350 model-pass outputs across three frontier LLM configurations, and further contextualized its outcomes through a matched direct-prompt baseline on a 150-file subset. The results show that \textsc{REFINE} substantially reduces detected code smells, especially major code smells, while broader quality indicators remain metric- and model-dependent and preservation risks persist. The contextual baseline suggests that \textsc{REFINE} provides more controlled candidate generation than direct prompting, with stronger overall code-smell reduction, smaller edits, and fewer public-method removals, although direct prompting preserves critical assert/fail constructs more often. Overall, \textsc{REFINE} offers a promising path toward traceable, evidence-aware LLM refactoring for targeted Java file-level code-smell mitigation, but generated candidates still require compilation, testing, dependency analysis, and human review before repository-level or production adoption.
\section*{Data Availability}

The \textsc{REFINE} implementation is available online~\cite{REFINEAno}, and the experimental dataset is available on Zenodo~\cite{REFINEDataset}.

\bibliographystyle{IEEEtran}
\bibliography{References}

\end{document}